\documentclass[amsmath,amssymb,twocolumn]{revtex4}

\usepackage[dvips]{graphicx}

\begin{document}

\title{Anti-localization of graphene under substrate electric field
}

\author{Ken-Ichiro Imura, Yoshio Kuramoto, Kentaro Nomura}
\affiliation{Department of Physics, Tohoku University, Sendai 980-8578, Japan}

\begin{abstract}
A simple criterion is provided how the (anti-)localization properties of graphene are determined in the presence of inter-valley scattering, Kane-Mele topological mass term, and Rashba spin-orbit interaction (SOI).
A set of (pseudo) time-reversal operations 
show that the number of effective internal degrees of freedom, such as
spin and pseudo-spins distinguishing the sublattice and the valley, is the crucial parameter for
localization.
It is predicted that perpendicular electric field due to gate voltage of the substrate drives the system to anti-localization by enhancing the Rashba SOI.

\end{abstract}


\maketitle

Graphene has a strong tendency 
{\it not to localize} \cite{SA,SA+,Nom07}, reflecting its linear spectrum \cite{Geim1}.
In ordinary two-dimensional metals, 
scaling either to weak localization (WL) or to weak anti-localization (AL) 
is controlled by impurity scattering with spin-orbit interaction \cite{HLN}.
If time reversal symmetry (TRS) is broken by external or internal magnetic field, 
the system exhibits neither WL nor AL
\cite{CSB}, and belongs to the unitary class
\cite{Dyson}.
Graphene shows the three localization classes even without magnetic impurities, but with potential scatterers.

\begin{figure}[b]
\includegraphics[width=8.5 cm]{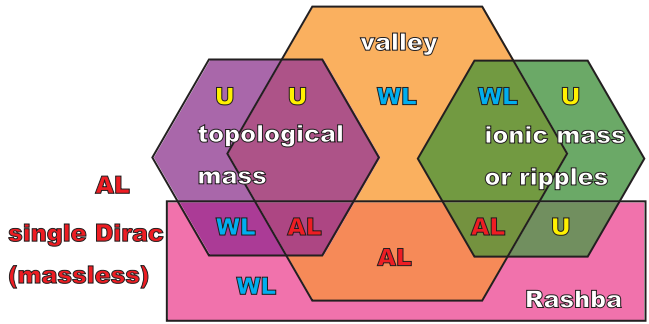}
\caption{
Weak localization classes of graphene-based models
with different types of the mass and impurity scattering.
WL, AL and U refer to weak localization (orthogonal class), 
weak anti-localization (symplectic class) and absence of WL
(unitary class), respectively.}
\label{venn}
\end{figure}

In the literature, theory predicts that graphene under doping
shows AL provided inter-valley scattering can be neglected
 \cite{SA,Nom07,SA+}.
Together with the absence of backward scattering,
AL in graphene is a clear manifestation of Berry phase $\pi$
\cite{ANS}. 
Inter-valley scattering, on the other hand, drives the system from AL to WL
\cite{SA}. 
Experiments \cite{Geim2,SiC} show also a unitary behavior.
Absence of WL may be attributed to ripples \cite{Guinea,Nom08}.
It is natural to ask how the localization properties are influenced by modification of the linear spectrum by a finite mass, whose magnitude is under debate in photoemission experiments \cite{Bostwick,DHLee}, and theoretically \cite{Min,KM}.
Another modification is due to
the Rashba SOI which inevitably appears in graphenes on substrates.
Kane and Mele have proposed that graphene with intrinsic spin-orbit interaction
realizes a $Z_2$ topological insulator
\cite{KM}. 
This paper demonstrates another striking aspect of graphene
under doping that crossovers between different symmetry class of
localization are controlled simply by the number of effective internal degrees of freedom.

The inter-valley coupling depends on the range of impurity potential \cite{NA}.
Long-range scatterers (LRS) do not involve intervalley scattering, 
whereas short-range scatterers (SRS) activates the valley spin. 
Localization properties of a disordered Kane-Mele model have been studied numerically \cite{MO,Obs}.  
Since tight-binding approximation is employed there, inter-valley scattering is always present, and cannot be controlled.
The strategy of our study is from the opposite direction;  we employ the lowest-order WL theory, but analyze systematically each element that influences the localization behavior. 

%
%

Our results are summarized in FIG.1,  which shows localization classes WL, AL, and U depending on the inter-valley scattering, types of the mass term, and the Rashba SOI.  Without these effects, the system becomes massless and belongs to AL, as is the case of single Dirac cone \cite{SA,Nom07}.
Note that, in the presence of inter-valley scattering, crossovers to AL occurs 
by switching on the Rashba SOI.  
The crossovers should be achieved by controlling the gate voltage of the substrate on which the system is placed.
In the single valley case, the Rashba SOI leads to either WL or U, but not to AL.
Such single valley system appears 
in HgTe/CdTe heterostructure \cite{Bernevig,HgTe}.

\begin{table*}[htdp]
\caption{Three story structure of Kane-Mele $Z_2$ topological insulator,
and its WL properties under doping.
LRS is equivalent to the single valley model.
The parity of $N_s$, the number of activated spin degrees of freedom, 
determines its WL properties: standard WL (orthogonal) or AL (symplectic).
Broken TRS leads to U (unitary) behavior.}
\begin{center}
\begin{tabular}{l|l|l}
\hline\hline
&
LRS (single valley)
&
SRS ($K$-$K'$ coupled)
\\ \hline \hline
(i) massless graphene: $H_1=p_x \sigma_x \tau_z+p_y \sigma_y$
&
$N_s=1$ ($AB$) $\rightarrow$ AL 
&
$N_s=2$ ($AB$,$KK'$)  $\rightarrow$ WL 
\\ \hline
(ii) mass terms: $H_2=H_1+H_{\Delta,m}$
&
unitary
&
(a) unitary
\\
(a) topological $-\Delta\sigma_z\tau_z s_z$ vs. (b) ionic $m\sigma_z$
&
no $1/g$-correction
&
(b) $N_s=2$ ($AB$,$KK'$) $\rightarrow$ WL 
\\ \hline
(iii) Rashba spin-orbit interaction:
&
(a) $N_s=2$ ($AB$, real spin) $\rightarrow$ WL 
&
$N_s=3$ ($AB$, $KK'$, real spin) $\rightarrow$ AL 
\\
$H_3=H_2-\lambda_R (\sigma_x \tau_z s_y-\sigma_y s_x)/2$
&
(b) unitary
&
\\
\hline\hline
\end{tabular}
\end{center}
\end{table*}

We take the doped and disordered Kane-Mele model for our weak-coupling perturbative theory.
The Kane-Mele model has a three story structure, depicted in TABLE I:
(i) graphene in the massless limit,
(ii) topological mass term, encoding Kane-Mele spin-orbit interaction,
(iii) Rashba term, playing the role of activating the real spin degree of freedom.
Note that the Kane-Mele model has also valley degree of freedom, corresponding to two Dirac points of graphene: $K$ and $K'$.
The ionic mass term is induced by chiral symmetry breaking staggered chemical potential. 
The Kane-Mele model possesses three types of pseudo or real spins, represented by
Pauli's matrices, $\vec{\sigma}$, $\vec{\tau}$ and $\vec{s}$,
operating in different subspaces:
$\vec{\sigma}$ acts on the sublattice spin $A$-$B$, $\vec{\tau}$ on the valley spin $K$-$K'$,
and $\vec{s}$ on the real spin.
In the continuum limit, the Kane-Mele Hamiltonian, 
\begin{equation}
H_{KM}=H_1+H_\Delta+H_R,
\end{equation}
consists of the following three elements:
(i) $H_1=\hbar v_F(p_x \sigma_x \tau_z+p_y \sigma_y)$,
(ii) $H_\Delta= -\Delta\sigma_z\tau_z s_z$,
(iii) $H_R= -\lambda_R (\sigma_x \tau_z s_y-\sigma_y s_x)/2$,
each describing the corresponding floor of the three story structure.
For comparison with $H_\Delta$, we consider also ionic mass term $H_m=m\sigma_z$.
In the ionic mass case, contributions to $\sigma_{xy}$ from the $K$- and $K'$-valleys 
cancel \cite{Sem}.
In the topological mass case, this cancellation of anomaly does not occur
any longer \cite{Hal}, 
and a quantized {\it spin} Hall effect with preserved TRS occurs.
In the absence of Rashba term $H_R$, the Hamiltonian $H_{KM}$ is {\it diagonal} 
in the real spin $\vec{s}$ space, implying that the latter is actually {\it inactive}.


As shown in TABLE I, sublattice spin $\vec{\sigma}$ is always active, whereas
valley and real spins can be switched on and off, leading to four different cases
for the subspace $\Sigma$ spanned by active spins.
In order to distinguish active and inactive spins, 
we introduce (pseudo) TRS operations $T_\Sigma$, 
defined in the subspace $\Sigma$, such that
\begin{equation}
T_\Sigma(H_1+H_R) T^{-1}_\Sigma 
= H_1+H_R,
\end{equation}
where
$\Sigma=\{\vec{\sigma}\}$, $\{\vec{\sigma}, \vec{\tau}\}$, $\{\vec{\sigma}, \vec{s}\}$,
$\{\vec{\sigma}, \vec{\tau}, \vec{s}\}$.  
Their explicit forms are given by
\begin{align}
&T_{\sigma}=-i\sigma_y C, \quad
T_{\sigma\tau}=\tau_x C, \nonumber\\
&T_{\sigma s}=(-i\sigma_y)(-is_y)C, \quad
T_{\sigma\tau s}=\tau_x (-is_y) C, \nonumber
\end{align}
where $C$ is complex conjugation.
$T_{\sigma\tau s}$ represents the genuine TRS operation.
Effective TRS of the system is, therefore, determined by the transformation property of
the mass term (see TABLE II).
When a mass term is odd against TRS, the system shows the unitary behavior.
Four unitary phases in FIG.1 correspond to the four minus signs in TABLE II.
If some (pseudo or genuine) TRS exists in the system,
its weak localization property is determined by the number $N_s$ of the activated spin degrees of freedom.
One can verify 
$T_\Sigma^2=1$ if $N_s$ is even, whereas
$T_\Sigma^2=-1$ if $N_s$ is odd.
The former (latter) corresponds to the orthogonal (symplectic) class
in the random matrix theory \cite{Dyson},
and leads to constructive (destructive) interference
between two scattering processes transformed from one to the other by $T_\Sigma$.
\begin{table}[b]
\caption{Time reversal operations $T_{\Sigma}$, 
relevant in the subspace spanned by activated spins.
Transformation property of a mass term 
${\cal O}=m\sigma_z,\ \Delta\sigma_z\tau_z s_z$
under $T_{\Sigma}$:
$
T_{\Sigma} {\cal O} T_{\Sigma}^{-1}
={\pm \cal O}$.
The sign appears in the table.
$U$ refers to unitary class.}
\begin{center}
\begin{tabular}{c|c|c|c|c}
\hline\hline
activated spins &$ \vec{\sigma}$ & $\vec{\sigma}, \vec{\tau}$ & $\vec{\sigma}, \vec{s}$ 
& $\vec{\sigma}, \vec{\tau}, \vec{s}$
\\ \hline
relevant TRS operation
&$T_{\sigma}$ & $T_{\sigma\tau}$ & $T_{\sigma s}$ & $T_{\sigma\tau s}$
\\ \hline \hline
$\sigma_z$ &$-\rightarrow U$&+&$-\rightarrow U$&+
\\ \hline 
$\Delta\sigma_z\tau_z s_z$ &$-\rightarrow U$&$-\rightarrow U$&+&+
\\ \hline \hline
\end{tabular}
\end{center}
\label{default}
\end{table}

In the presence of a mass term, irrespective of its type, 
one finds unitary behavior for LRS.
Scattering matrix elements are {\it diagonal} in $\vec{\tau}$-space, 
and $\vec{\sigma}$ is the only active spin ($N_s=1$, TRS broken).
In the massless case, a pseudo TRS operation $T_{\sigma}$ mimics the role of genuine TRS \cite{Lud}.
Once TRS is effectively restored, the system's WL property is determined by the parity of $N_s$.
The mass term, on the other hand, explicitly breaks $T_{\sigma}$.

SRS activate the valley spin $\vec{\tau}$ ($N_s=2$), since its matrix elements
involve off-diagonal terms in this subspace.
In the case of ionic mass term, this leads the system to standard WL,
since activation of the valley spin $\vec{\tau}$ restores the (pseudo) TRS.
In the case of topological mass term, the system stays unitary,
since restoration of TRS needs also the activation of real spin $\vec{s}$.
The latter is embodied by the Rashba SOI.
In the presence of both SRS and Rashba SOI, we predict AL, since $N_s=3$.

Starting with the case of $\lambda_R=0$,
let us go into some details of diagrammatic calculations. 
In the presence of topological mass term $H_\Delta$,
The spinor part of the conduction band eigenstates reads,
\begin{equation}
|K\alpha\rangle =
\left[
\begin{array}{c}
 \cos{\theta_\alpha\over 2} \\
e^{i\phi_\alpha}\sin{\theta_\alpha\over 2}
\end{array}
\right],
|K'\alpha\rangle =
\left[
\begin{array}{c}
e^{i\phi_\alpha}\sin{\theta_\alpha\over 2} \\
- \cos{\theta_\alpha\over 2}
\end{array}
\right],
\label{spinor2}
\end{equation}
where $\alpha$ specifies a three-dimensional fictitious momentum 
$\vec{p}=(p_x, p_y, -\Delta)$,
and the polar angles $\theta$, $\phi$ satisfy
$\cos\theta=-\Delta/ \sqrt{p_x^2+p_y^2+\Delta^2}$,
$\cos\phi=p_x/\sqrt{p_x^2+p_y^2}$.

LRS have a potential range much larger than the inter-atomic distance, 
and do not couple $K$ and $K'$.
The system cannot see the difference between two types of mass term,
both showing unitary behavior, i.e.,
the diffusion type singularity is cut off by a cooperon's lifetime. 
This unitary phase shows a crossover to the well-established symplectic behavior 
of graphene in the single Dirac cone \cite{SA,Nom07}.

\begin{figure}[!]
\includegraphics[width=8.5 cm]{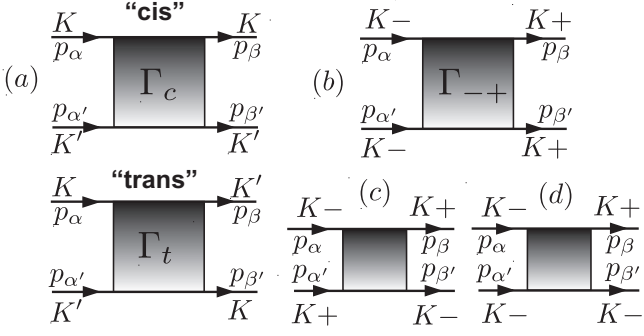}
\caption{Particle-particle ladders. 
(a) Relevant diagrams in the presence of short-range scatterers (SRS).
"cis" and "trans" refers to specific configurations of the valleys.
(b-d) Bare diagrams involving {\it inter-branch} processes 
($\lambda_R \ne 0$, no SRS).
(b) $\gamma_{-+}$ contributes to $1/q^2$-singularity, 
whereas such diagrams as (c) and (d) are irrelevant to the singularity,
since $p_\alpha+p_{\alpha'}$ cannot be smaller than the order of
$\lambda_R$.}
\label{ladders}
\end{figure}

SRS involve inter-valley scattering, allowing for
distinguishing the two different types of mass term:
topological and ionic.
The scattering matrix elements involve a projection operator in the AB sublattice space,
${\cal P}_{A,B}$.
As for the singular contribution, one can focus on the two types of diagrams 
shown in FIG.1.
The ``trans" component $\gamma_{t}$ reads explicitly, 
\begin{eqnarray}
\gamma_{t} &=&
2\pi\nu n_A u_A^2
\langle K'\beta|{\cal P}_A \tau_-|K\alpha\rangle
\langle K\beta'|{\cal P}_A \tau_+|K'\alpha'\rangle
\nonumber \\
&+&2\pi\nu n_B u_B^2
\langle K'\beta|{\cal P}_B \tau_-|K\alpha\rangle
\langle K\beta'|{\cal P}_B \tau_+|K'\alpha'\rangle
\nonumber \\
&=& - e^{i(\phi_\alpha-\phi_\beta)}
\eta_S \sin^2 \theta /2
\ (=-\gamma_{c}),
\label{trans}
\end{eqnarray}
where we have introduced
$\eta_S=2\pi\nu (n_A u_A^2+n_B u_B^2)/2$,
with $\nu$, $n_{A,B}$ and $u_{A,B}$ being, respectively, the density of states,
the impurity density and the typical strength of scattering potential 
at the A (B) sites.
$\gamma_t$ has an additional minus sign,
which plays the role of driving the crossover from symplectic to orthogonal behavior
in the massless limit \cite{SA}.
The Bethe-Salpeter equation (BSE) takes the form of two coupled equations:
\begin{equation}
\left[
\begin{array}{c}
\Gamma_{c}\\
\Gamma_{t}
\end{array}
\right]_{\alpha\beta}
=\left[
\begin{array}{c}
\gamma_{c}\\
\gamma_{t}
\end{array}
\right]_{\alpha\beta}
+\left[
\begin{array}{cc}
\gamma_{c}&\gamma_{t}\\
\gamma_{t}&\gamma_{c}
\end{array}
\right]_{\alpha\mu}
\Pi_\mu
\left[
\begin{array}{c}
\Gamma_{c}\\
\Gamma_{t}
\end{array}
\right]_{\mu\beta}
\label{BSE_ct}
\end{equation}
After diagonalization, one finds,
$\Gamma_{c} +\Gamma_{t}=0$,
and
$\left[1-(\gamma^{(1)}_c -\gamma^{(1)}_t)\Pi_S \right](\Gamma_{c} -\Gamma_{t})=
\gamma_{c} -\gamma_{t}$,
where  $\Pi_S\simeq \tau_S(1-\tau_S Dq^2)$ with $\tau_S=1/\eta_S$ being the scattering time.
We also introduced $\gamma^{(1)}_{c,t}$ in the light of 
a general expression: $\gamma=\sum_l \gamma^{(l)} e^{i l(\phi_\alpha-\phi_\beta)}$.
Cancellation between $\tau_S$ and the bare $\gamma$'s 
is incomplete, giving a finite lifetime.
The system shows a unitary behavior, irrespective of
the preserved TRS of underlying Hamiltonian $H=H_1+H_\Delta$.
$1/q^2$-singularity is recovered in the limit $E\rightarrow \infty$
(orthogonal class).

In the case of ionic mass term, a complete cancellation between 
the self-energy and the bare vertex function occurs, driving 
a crossover from unitary to orthogonal symmetry class.
This situation is relevant to the intrinsic single valley system \cite{Bernevig,HgTe}
in the presence of off-diagonal interaction.

Rashba SOI appears
when inversion symmetry with respect to the 2D plane is broken,
say, by a perpendicular electric field.
Rashba SOI lifts the two-fold real spin degeneracy 
of the two conduction and two valence bands.
An accidental degeneracy occurs on top of the valence bands,
whereas the conduction bands $E_{u\pm}$ are split by $2\lambda_R$:
$E_{\pm}=\sqrt{p_x^2+p_y^2+(\Delta\pm\lambda_R/2)^2}\pm\lambda_R/2$.
In order to parametrize the corresponding eigen spinors, it is convenient to
introduce fictitious 3D momenta:
$\vec{p}_\pm=(p_x, p_y, \Delta\pm\lambda_R/2)$, 
and 
$\cos\theta_\pm=(\Delta\pm\lambda_R/2)/\sqrt{p_x^2+p_y^2+(\Delta\pm\lambda_R/2)^2}$.
In terms of these parameters,
the eigen spinors corresponding to $E_{\pm}$ read,
\begin{equation}
|K\pm \rangle \propto \left[
\begin{array}{c}
e^{-i\phi}\sin{\theta_\pm\over 2}\\
 \cos{\theta_\pm\over 2}\\
\mp i \cos{\theta_\pm\over 2}\\
\mp i e^{i\phi} \sin{\theta_\pm\over 2}
\end{array}
\right],
|K'\pm \rangle \propto
\left[
\begin{array}{c}
\cos{\theta_\pm \over 2}\\
- e^{-i\phi}\sin{\theta_\pm \over 2}\\
\mp i e^{i\phi} \sin{\theta_\pm \over 2}\\
\pm i \cos{\theta_\pm \over 2}
\end{array}
\right]
\label{spinor4}
\end{equation}
where we omitted the normalization factor $1/\sqrt{2}$.
Rashba coupling imposes a stronger constraint on the choice of our basis.
As a result, the matrix element, such as,
$\langle K\beta| 1 |K\alpha\rangle = 
\cos (\phi_\alpha-\phi_\beta) \sin^2 (\theta_- /2)+
\cos^2 (\theta_- /2)$,
becomes {\it real} (no Berry phase).

When $E>\Delta+\lambda_R$,
inter-branch matrix elements plays a role.
They modify the scattering time $\tau_\pm$ for the $|Ku \pm \rangle$ branches
as,
\begin{equation}
{1\over\tau_\pm}={\eta_L\over 2}\left[
\sin^4{\theta_\pm\over 2} + 2 \cos^4{\theta_\pm\over 2} +
\sin^2{\theta_-\over 2}\sin^2{\theta_+\over 2}
\right].
\label{tau+-}
\end{equation}
As for particle-particle ladders, such as FIG.2 (b),
four electron states $\alpha$, $\beta$, $\alpha'$, $\beta'$ 
can, in principle, take either of the two channel indices, $|K- \rangle$ or $|K+ \rangle$,
generating eight types of diagrams in total.
However, a simplification is possible at this level:
since we are interested only in the $1/q^2$-singular part of cooperon diagrams,
we need $p_\alpha+p_{\alpha'}=q\simeq 0$.
This means that $\alpha$ and $\alpha'$ must belong to the same branch.
We can thus safely focus on such diagrams as
$\gamma_{+-}$ and $\gamma_{-+}$ depicted in FIG.2 (b),
or even simpler
$\gamma_{--}$ and $\gamma_{++}$.
Other $\gamma$'s such as FIG.2 (c) and (d) are irrelevant to $1/q^2$
singularity.
Explicit form of relevant $\gamma$'s are given as,
\begin{eqnarray}
\gamma_{\pm\pm}&=&  \eta_L \Big[ \cos^2 (\phi_\alpha-\phi_\beta) \sin^4 {\theta_\pm \over 2} 
\nonumber \\
&+& 2\cos^2 (\phi_\alpha-\phi_\beta) \sin^2 {\theta_\pm \over 2} \cos^2 {\theta_\pm \over 2}+
\cos^4 {\theta_\pm \over 2} \Big]
\nonumber \\
\gamma_{\pm\mp}&=& - \eta_L \sin^2 (\phi_\alpha-\phi_\beta) \sin {\theta_+\over 2} \sin {\theta_-\over 2}.
\end{eqnarray}
These four types of diagrams satisfy coupled BSE.
But its $4\times 4$ coupling matrix is shown to be block diagonalizable.
One can, e.g., decouple $\Gamma_{--}$ and $\Gamma_{+-}$, from the remaining part.
The former two obey coupled equations, which take symbolically the following form:
\begin{equation}
\left[
\begin{array}{c}
\Gamma_{--}\\
\Gamma_{+-}
\end{array}
\right]
=\left[
\begin{array}{c}
\gamma_{--}\\
\gamma_{+-}
\end{array}
\right]
+\left[
\begin{array}{cc}
\gamma_{--}\Pi_-  &\gamma_{-+}\Pi_+ \\
\gamma_{+-}\Pi_- &\gamma_{++}\Pi_+
\end{array}
\right]
\left[
\begin{array}{c}
\Gamma_{--}\\
\Gamma_{+-}
\end{array}
\right],
\label{BSE+-}
\end{equation}
where  $\Pi_\pm\simeq \tau_\pm (1-\tau_\pm Dq^2)$ with
$\tau_\pm$ given in Eq.(\ref{tau+-}).
To identify the singular contribution, 
we employ the expansion into different angular momentum contributions, 
both for $\gamma$'s and $\Gamma$'s as
$\Gamma=\sum_l \Gamma^{(l)} e^{i l(\phi_\alpha-\phi_\beta)}$,
and pick up only the $l=0$ component. 
The following identity,
\begin{equation}
\det
\left[
\begin{array}{cc}
1-\gamma^{(0)}_{--}\tau_-  &- \gamma^{(0)}_{-+}\tau_+\\
-\gamma^{(0)}_{+-}\tau_-  &1-\gamma^{(0)}_{++}\tau_+
\end{array}
\right]=0,
\label{det0}
\end{equation}
ensures that the dressed cooperons show indeed $1/q^2$-singularity at the $l=0$ channel.
Rashba SOI thus drives the system to standard WL,
whenever the Fermi level is above the gap.

If one adiabatically switches {\it off} the Rashba term,
the simplification we have made for justifying Eq.(\ref{BSE+-})
is no longer valid.
In the limit of vanishing $\lambda_R$ we cannot simply neglect such
diagrams as FIG.1 (c), (d).
They could contribute equally to the $1/q^2$ singularity
if the singularity ever appears.
Relations such as $p_\alpha+p_{\alpha'}=q\simeq 0$ can be satisfied
in these diagrams.
In the coupled BSE, the cooperons acquire more channels to couple to,
and one can no longer decouple $\Gamma_{--}$ and $\Gamma_{+-}$
as Eq.(\ref{BSE+-}).
As a result, the cancellation property between particle-particle ladders and 
the self-energy such as Eq.(\ref{det0}) is lost.
The loss of cancellation property immediately leads to the absence of WL.
On the contrary, if one starts from the situation where Rashba SOI is absent in the first place,
it is possible to choose a basis in which that two (degenerate) channels are decoupled.

As for the ionic mass, Rashba SOI plays no role and the system stays unitary.
The same applies to the case of ripples \cite{Nom08}, 
the latter possessing the same symmetry properties as the ionic mass.

We finally consider the most realistic case of SRS in the presence of Rashba SOI ( $\lambda_R\ne 0$). 
One has to use (\ref{spinor4}) instead of
(\ref{spinor2}), and develop a similar analysis. 
One finds in this case the {\it cancellation} between the self-energy and $\gamma$
at the $l=0$ channel.
But here additional minus sign analogous to Eq.(\ref{trans}) appears
due to inter-valley scattering, driving the system to AL.

Let us estimate the strength of gate electric field
required for observing the crossover to AL.
For the crossover to be experimentally accessible,
Rashba SOI needs to be the order of $\sim 1$ K.
This corresponds to the electric field of order $\sim 1 $V/nm \cite{Min},
a value attainable in double-gated graphene devices \cite{Vanders}.
The crossover to AL will be observed for a sample with insignificant ripples.  
A similar crossover due to Rashba SOI has been observed in another context
in InGaAs/InAlAs quantum well \cite{Nitta}.
In the case of graphene, 
crossovers occur not only from WL, but also from the unitary class \cite{Geim2}.

In conclusion, we have identified key elements responsible for rich localization properties of graphene that are controllable by substrate electric field.  The criterion whether the system is driven to WL or AL is simply given by the number of active (pseudo-)spin degrees of freedom.



\end{document}